\input amstex

\documentstyle{amsppt} \magnification1200 \TagsOnRight 
\NoBlackBoxes
\pagewidth{30pc} \pageheight{47pc} 

\input epsf.tex


\def\today{\number\day\space\ifcase\month\or January\or February\or
March\or April\or May\or June\or July\or August\or September\or
October\or November\or December\fi\space\number\year}
 
\def\SL{\operatorname {SL}} 
\def\SU{\operatorname {SU}} 
\def\su{\operatorname {su}} 
\def\SO{\operatorname {SO}}
\def\so{\operatorname {so}}
\def\Tr{\operatorname {Tr}}
\def\C{{\Bbb C}}

\def\R{{\Bbb R}}

\topmatter
\title  Relativistic spin networks \\
and quantum gravity\endtitle
\author John W. Barrett\\ Louis Crane\endauthor
\date 12 September 1997; revised 21 January 1998\enddate     
\address
Department of Mathematics,
University of Nottingham,
University Park,
Nottingham,
NG7 2RD, UK
\endaddress
\email jwb\@maths.nott.ac.uk \endemail
\address
 Mathematics Department, 
Kansas State University, 
Manhattan KS, 66502, USA
\endaddress
\email crane\@math.ksu.edu\endemail

\abstract
Relativistic spin networks are defined by considering the spin covering of the group $\SO(4)$, $\SU(2)\times \SU(2)$. Relativistic quantum spins are related to the geometry of the 2-dimensional faces of a 4-simplex. This extends the idea of Ponzano and Regge that $\SU(2)$ spins are related to the geometry of the edges of a 3-simplex. This leads us to suggest that there may be a 4-dimensional state sum model for quantum gravity based on relativistic spin networks which parallels the construction of 3-dimensional quantum gravity from ordinary spin networks. 
 
\endabstract

 \endtopmatter
 
\newpage

\document

\head I. Introduction \endhead 

In \cite{1} and \cite{2}, it was proposed that the quantum theory of gravity
could take the form of a very special type of discrete model on a
triangulated 4-manifold, which has the property that the propagation
of physical states does not depend on the choice of triangulation.
Such a model is called a topological state sum.

Topological state sums are closely related to topological quantum field theories. Arguments that the quantum theory of gravity should be closely related
to a topological quantum field theory are given in \cite{3}.

In \cite{4} a specific family of four dimensional topological state sums were constructed.
It was suggested in \cite{1} and \cite{2} that the topological state sum of \cite{4} might be a
representation of a topological state for the quantum theory of gravity,
and that the physical hypothesis that the universe is in that state
could be studied.

Ooguri \cite{5}, and more recently, Baez \cite{6} have argued that the topological 
state sum of \cite{4} (reformulated in \cite{7}), which is a precise realisation of \cite{5}, is a construction of the quantum theory of the $BF$ lagrangian.   
 
In fact, the Plebanski Lagrangian \cite{8,9} for general relativity has
the form of $BF$ theory \cite{10} with gauge group $\SL(2,C)$, with an added term which is a constraint.   Smolin \cite{11} has
argued that for general relativity with a cosmological constant the
constraint vanishes for the self-dual sector.  This suggests \cite{1,2} that the self-dual sector can be identified with the topological state sum of \cite{4}.

The purpose of this paper is to outline a new state sum model which is a sort of
constrained doubling of the topological state sum of \cite{4}.  This constraint is a quantum version of the Plebanski constraint.

In three dimensions, quantum gravity is constructed by the Ponzano-Regge state sum model in which representations of $\SU(2)$ label edges \cite{12}. This corresponds to a path integral in which the metric is the variable. However this model can also be understood as a path integral in the first-order formalism  using a frame field and connection form.\cite{13} 

It is worth pursuing the link with the classical frame field to get a heuristic understanding of the discrete models. Integrating the frame field $e$, an $\R^3$-valued 1-form, along an edge gives a vector in $\R^3$. This can also be regarded as an element of the dual of the Lie algebra $\so(3)$.

The idea is that to pass from 3 dimensions to 4 dimensions one has to replace   $\SO(3)$ with $\SO(4)$, or its spin covering, $\SU(2)\times\SU(2)$.

The topological home for this spin is now on triangles. This accords with the assignment of spins to triangles in \cite{4}, which can be regarded as a special case of $\SU(2)\times\SU(2)$, where one of the copies of $\SU(2)$ acts trivially, thus rendering all variables self-dual.  We are also making contact with the loop variables
picture of Rovelli and Smolin \cite{14}, in which the fundamental operators
live on surfaces. The relation of the loop variables to state sum models, and the difference between $BF$ theory and gravity in this context,  
was investigated in \cite{15} and \cite{16}.

In a $BF$ theory, the $B$ field is an $\so(4)^*$-valued 2-form, which can be integrated on a triangle to give an element of $\so(4)^*$, the dual of the Lie algebra $\so(4)$. However, in Einstein gravity it is important that $\so(4)^*$ is identified with $\R^4\wedge\R^4$, as $B$ is constrained to be of the form $ B=*e\wedge e$, where $e$ is a vector valued one-form. In this formula, $e\wedge e$ is an $\R^4\wedge\R^4$-valued 2-form, and $*$ is the Hodge $*\colon\R^4\wedge\R^4\to\R^4\wedge\R^4$.
 
Assuming the frame field gives a linear embedding of the triangle in $\R^4$, the integral over this triangle,
$$\int e\wedge e$$
is a {\it simple} bivector, i.e., of the form $f\wedge g\in \R^4\wedge\R^4$. Thus the essential constraint which marks Einstein gravity from $BF$ theory is the simplicity of bivectors. 

This remark is the key to our construction of the discrete model, even though we do not have a classical frame field in the construction.  The discrete models and the path integral formalism with classical frame fields do have close parallels, however.

A discretisation of the classical Einstein gravity using  bivectors on triangles which are constrained to be simple was previously studied in \cite{17,18}. Some related discrete classical models with self-dual bivectors, rather than simple bivectors, were studied in \cite{18,19}.

The new state sum model gives a quantum mechanical version of these constraints. The translation is to replace bivectors with bivector operators. The simplicity condition is then translated into a condition in the representation theory for these operators.

At present we have an axiomatic description of the space of vertices but not an explicit construction. If the only allowable vertex is the one we explicitly
construct, then the state sum does not seem to give a topological
invariant of manifolds. Thus the relationship of our new state sum to topology is still unclear.

However, it is already very close to a topological theory, in that the deviation from the topological theory \cite{4} is only in the imposed constraints.
This should make investigation of its `ultraviolet' behavior much simpler
than for a continuum theory. Our hope is that the `doubling' of
the topological state sum of \cite{4} should allow us to pass from the self-dual sector to the full quantum theory of gravity.

The new state sum has a very natural relationship with Regge calculus in a
new formulation, in that the classical version of the theory has the metric  as its variable in the same way as Regge calculus.  We believe, though do not have a complete argument, that it will contain `local degrees of freedom' as does general relativity and that propagating plane waves
will occur in a linearization of the theory as in \cite{20,21}. However we do not yet have an analogue of the Ponzano-Regge asymptotic formula which would give a direct link with the Einstein action.

The heart of our construction is a reformulation of the geometry of
simplexes in Euclidean (or Minkowski) 4-space in terms of variables on their 2-subsimplexes.  This paper, accordingly, begins with a theorem in classical
4-dimensional Euclidean geometry.  The theorem allows
us to describe 4-simplexes in terms of oriented area variables on
2-simplexes, giving a more precise version of the idea in \cite{17}.  The constraints on the variables have a very elegant
form, which can be carried over to the quantum theory.

Having discovered this form for the quantum constraints, it then
becomes relatively straightforward to produce the new state sum.  
   
\head II. Bivector geometry of a 4-simplex\endhead
An abstract 4-simplex is determined by an ordered set of 5 points, which we denote $(01234)$. The faces of the abstract 4-simplex are just the subsets of this set.

A geometric 4-simplex is the convex hull of 5 points in $\R^4$.  This embedding is required to be non-degenerate, i.e., the 5 points do not lie in any hyperplane. To each face now corresponds an affine subspace in $\R^4$, and also a linear subspace through the origin in $\R^4$ which is parallel to it. In particular, each triangle (2-simplex) determines a plane through the origin.

An embedding of an abstract 4-simplex in $\R^4$ is simply the set of position coordinates $x_0, x_1, x_2, x_3, x_4 \in\R^4$ which determine the vertices of a geometric 4-simplex. The ordering of the vertices also determines an orientation for the simplex.

An oriented triangle in $\R^4$ determines a bivector. The position coordinates $x_0, x_1, x_2 \in\R^4$ determine the displacement vectors for the edges $e_{01}=x_0-x_1$, $e_{02}=x_0-x_2$ and hence the bivector $b={1\over2}e_{01}\wedge e_{02}\in \R^4\wedge\R^4$. This bivector is invariant under an even permutation of the vertices but changes sign for an odd permutation.
 
In this way, a geometric 4-simplex determines a set of 10 bivectors, one for each triangle. We show that the bivectors characterise the geometric 4-simplex uniquely, and give the conditions that an assignment of 10 bivectors to the triangles of an abstract 4-simplex determines an embedding.

The properties of the bivectors for a 4-simplex are

\roster\item The bivector changes sign if the orientation of the triangle is changed.
\item Each bivector is simple, i.e. of the form $b=f\wedge g$.
\item If two triangles share a common edge, then the sum of the two bivectors is also simple.
\item The sum of the 4 bivectors corresponding to the faces of a tetrahedron is zero. This sum is calculated using the orientation of the triangles given by the boundary of the tetrahedron.
\item The assignment of bivectors is non-degenerate. This means that for six triangles sharing a common vertex, the six bivectors are linearly independent.
\item Using the Euclidean metric, the bivectors can be considered as linear operators. Then for 3 triangles meeting at a vertex of a tetrahedron, $\Tr b_1[b_2,b_3]>0$. The sign in this formula is determined by a convention relating a choice for the orientation for the boundary of the tetrahedron to the order in which the three bivectors appear in this formula.
\endroster
 Conditions \therosteritem2, \therosteritem3 and \therosteritem4 were introduced in \cite{17}. We are indebted to John Baez \cite{22} who pointed out the necessity of \therosteritem6 and the inversion
ambiguity in the theorem which were missing in our first version of this paper.

\proclaim{Theorem} Each geometric 4-simplex determines a set of bivectors satisfying these conditions, and each set of bivectors satisfying these conditions determines a geometric 4-simplex unique up to parallel translation and inversion through the origin.
\endproclaim

\demo{Proof} A simple bivector determines a plane (of dimension 2) through the origin in $\R^4$. Two simple bivectors $b_1$, $b_2$ determine two planes lying in a hyperplane (of dimension 3) if and only if $b_1+b_2$ is simple. If the planes lie in a hyperplane, this is clear because all bivectors are simple in three dimensions. Conversely, the simplicity condition can be understood as $b\wedge b=0$. Then
$$0=(b_1+b_2)\wedge(b_1+b_2)=2 b_1\wedge b_2.$$
If $b_1=x\wedge y$, $b_2=z\wedge t$, then $0=x\wedge y\wedge z\wedge t$ implies a linear relation between $x,y,z,t$. This calculation also shows that \therosteritem3 is insensitive to the orientation of the triangles. 

For the bivectors for a geometric 4-simplex \therosteritem4 is satisfied by Stokes' theorem. Condition \therosteritem5 is true because the edges meeting at a vertex can be taken as axes for $\R^4$, and then the bivectors are clearly linearly independent. Condition \therosteritem6 follows from the calculation $\Tr b_1[b_2,b_3]={9\over8}V^2$, where $V$ is the volume of the tetrahedron \cite{23}. This completes the proof that the conditions are satisfied by a geometric 4-simplex.

Now suppose a set of bivectors $b_i$ satisfies  \therosteritem1--\therosteritem6. 
The simplicity conditions and the non-degeneracy imply that the planes corresponding to the four faces of a tetrahedron either all lie in a common hyperplane or share a common direction.  However if the bivectors share a common direction, then $\Tr b_1[b_2,b_3]=0$, contradicting \therosteritem6. Therefore the four planes for a tetrahedron lie in a hyperplane. Furthermore, each plane is determined as the intersection of two of these hyperplanes, as each triangle is contained in two tetrahedra.
The non-degeneracy condition implies that the five hyperplanes intersect generically, that is, any four of them intersect at a point. 
 
To construct the geometric 4-simplex, shift one of the five hyperplanes away from the origin by parallel translation. The hyperplanes now bound a geometric 4-simplex, with 2-dimensional faces parallel to the planes corresponding to the original bivectors $b_i$, $i=1,\ldots10$. This simplex has 10 bivectors $b_i'=\lambda_i b_i$ for some 10 scalars $\lambda_i\ne0\in \C$. Condition \therosteritem4 is the only linear relation on the bivectors for the faces of a tetrahedron. This implies that the $\lambda_i$ are all equal for the faces of a tetrahedron, and hence are all equal for all ten triangles.

The geometric 4-simplex can be scaled by moving the hyperplane. If its distance from the origin is scaled by a parameter $\mu$, then the bivectors all scale as $\mu^2$. Therefore the  $\lambda_i$ can be scaled so that they are all equal to $1$ or all equal to $-1$.  Then the area bivectors $b_i'$ of the reconstructed 4-simplex are either exactly equal to the original $b_i$, or all equal to $-b_i$. However the $b_i'$ also satisfy \therosteritem6 as they are the bivectors of a geometric 4-simplex.
Hence the inequality in \therosteritem6  forces the $b_i'$ to be exactly equal to the $b_i$ and not to their negatives. 

The scaling parameter $\mu$ is only determined up to a sign, and so the geometric 4-simplex is determined up to inversion about the origin. This operation does not affect the area bivectors.  \enddemo

\subhead Metric geometry \endsubhead
Now introduce the standard Euclidean metric on $\R^4$ (One could also use the standard Minkowski metric).  If the bivectors are replaced by their duals, $*b$, then conditions \therosteritem1 -- \therosteritem5 are still satisfied but  $\Tr b_1[b_2,b_3]=0$. This duality interchanges a set of planes lying in a hyperplane with a set of planes sharing a common direction.
 
The bivectors can be identified with elements of the Lie algebra $\so(4)$. (Also, it may be profitable to think of them as the elements of the dual of this Lie algebra.)
The splitting $\so(4)\simeq \su(2)\oplus\su(2)$ is then the same as the splitting of the space of bivectors into self-dual and anti-self-dual parts,
$\R^4\wedge\R^4=\Lambda^2_+\oplus\Lambda^2_-$. The condition that a bivector $b$ is simple is
$$0=<b,*b>=<b^+,b^+>-<b^-,b^->,$$
 so that the norm of the self-dual and anti-self-dual parts is equal. The area $A$ of the triangle is determined by
$$A^2=<b,b>=<b^+,b^+>+<b^-,b^->.$$

A geometric simplex has a metric geometry, which is determined by the lengths of its 10 edges. Thus two geometric 4-simplexes have the same edge lengths if and only if they are related by an isometry of $\R^4$. Therefore the bivector data determines a set of edge lengths. Two sets of bivector data determine the same edge lengths if and only if the bivectors are all transformed simultaneously by an orthogonal rotation.

By contrast, specifying the 10 areas does not specify the metric geometry uniquely \cite{24}, although it does for almost all configurations. However giving the areas of the four faces of a tetrahedron does not specify the geometry of the tetrahedron uniquely, so that two 4-simplexes cannot be glued together by specifying the areas of the 16 triangles in their union \cite {25}. For this reason, we decided that the theory should be based on bivectors rather than areas.

\head III. Relativistic spin networks\endhead

In the theory of spin networks, using the Lie group $\SU(2)$, each representation determines a map from vectors in $\R^3$ to operators in the representation space, namely the representation of the Lie algebra. In quantum physics, these are called vector operators. The intuition is that in the classical limit, when the representation is `large', these behave as ordinary vectors in 3-dimensional space, with probability near to 1.

Given three irreducible representations on spaces $V_a$, $V_b$, $V_c$, the spin network trivalent vertex is a uniquely determined element of
$$ V_a \otimes V_b\otimes V_c.$$ In a three-dimensional theory, this is the vertex for a triangle. This element is invariant under the group action. The corresponding vector operators satisfy
$$ J_a+J_b+J_c=0$$
when applied to this state, as a consequence. This is the relation for the edge vectors of a triangle, or, equivalently, the bivectors in $\R^3$ which are dual to the edge vectors. Note that for other groups, the space of vertices may form a vector space of dimension greater than one. In these cases, it is necessary to pick a basis for these spaces and sum over the basis in a state sum model.

To define a theory of relativistic spin networks, we label the edges of a graph with irreducible representations of $\SU(2)\times\SU(2)$. These are pairs $(j,k)$ of representations of $\SU(2)$.
Then the representation of the Lie algebra gives bivector operators, by analogy with the preceding case. The condition that these are simple bivectors can be transcribed into the quantum theory by requiring that $j=k$. This is the correct condition because classically it is requiring the self-dual and anti-self-dual parts to have the same norm.

We also have in mind using the $q$-deformation of spin networks \cite{26}. This is algebraically very similiar to the undeformed case, with the advantage that the computations can be made finite by the choice of a root of unity. A disadvantage is the loss of the simple geometrical picture of bivectors. Also, the representations are braided. We take the braiding for the two copies of $\SU(2)$ to be opposite, i.e., related by an orientation reversing map.

Now in the relativistic case, the theory is for four-dimensional manifolds and the relativistic spin network vertices reside in a tetrahedron. There will be one bivector operator for each face of a tetrahedron, and so this will be a four-valent vertex.
The idea for relating spins on faces to the geometry of a tetrahedron is from Barbieri \cite{23}, who considered a single $\SU(2)$ spin.

The labelling of a 4-simplex is as follows.  The triangles are labelled with representations of $\SU(2)\times\SU(2)$, and each tetrahedron is labelled with a tensor in the product of the four spaces on its faces.
\roster\item Changing the orientation of a triangle changes the representation to its dual. \item The representations on the triangles are of the form $(j,j)$. These are called simple representations.
\item For any pair of faces of a tetrahedron, the pair of representations can be decomposed into its Clebsch-Gordan series for $\SU(2)\times\SU(2)$. Under this isomorphism, the tensor for the tetrahedron decomposes into summands which are non-zero only for the simple representations of $\SU(2)\times\SU(2)$.
\item The tensor for the tetrahedron is an invariant tensor. 
\endroster
These conditions are the quantum analogues of the properties \therosteritem1 -- \therosteritem4 of bivectors on a 4-simplex given earlier. For the first condition, note that in the case of a classical Lie algebra, the action of the Lie algebra in the dual representation is minus the adjoint of the original action. We have not analysed the inequalities \therosteritem5 and \therosteritem6, but note that a simple operator analogue of \therosteritem6 seems feasible. Operator versions of the expression in \therosteritem6 are considered in the context of canonical quantization in \cite{27}. 
 
We do not have a description of a basis for the space of vertices. However we can show that the space is non-empty in many cases, by exhibiting a canonical element. This vertex is defined by
$$\sum_j \dim j\quad \vcenter{\epsfbox{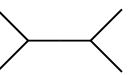}}$$
where the middle edge is labelled by $(j,j)$, and $\vcenter{\epsfbox{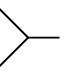}}$ is the usual $\SU(2)$ vertex doubled.
 
This looks unsymmetric because the faces of the tetrahedron have been split into two pairs. Each term in the sum obviously satisfies \therosteritem 2 and \therosteritem4, and satisfies \therosteritem3 for the two preferred pairs. The sum of the bivector operators for a pair is equal to the bivector operator for the middle edge in the vertex. This is a simple bivector operator because only simple representations of the form $(j,j)$ appear. 

However the pairs can be switched by fusion, with the coefficients determined by the 6j-symbol. The orthogonality relation of the 6j-symbol shows that the vertex is given by the same formula if the top two and bottom two pairs are chosen. For the third splitting, the braidings of self dual and anti self dual pieces cancel, so the same remark applies.
 
\subhead The classical limit \endsubhead
The simple representations of  $\SU(2)\times\SU(2)$ occur as the summands of the Hilbert space of functions on $S^3$, which has the obvious action of $\SO(4)$. The corresponding classical phase space is therefore the cotangent space $T^*S^3$. A Lie algebra element $l\in\so(4)$ acts on functions on $S^3$ by the formula
$$l=l^{\mu\nu}x_\mu{\partial\over\partial x^\nu}.$$
The corresponding momentum in the classical case is $l^{\mu\nu}x_\mu p_\nu$, for a point $(x,p)\in T^*S^3$. This is the dual pairing of $\so(4)$ with the simple bivector $x\wedge p$ in $\R^4$, which explains how simple bivectors are associated to triangles. 

This phenomenon is familiar from the standard relativistic angular momentum theory, where the failure of simplicity of the angular momentum is associated with intrinsic spin. In our picture, intrinsic spin occurs for non-simple representations, which would arise from sections of spinor bundles over $S^3$.

\head  IV. The New Model\endhead

It is easy to see how to transform the state sum
of \cite{4} to create a quantum version of summation over all assignments of
bivector data to the 2-simplexes of a triangulated 4-manifold.  We pick
labellings of the 2-simplex of the triangulation by
irreducible representations of $U_q(SU_2)$ just as in \cite{4}. For each such labelling of the 2-simplexes, we label the tetrahedra in each 4-simplex by elements of a basis for the vector space of vertices defined in the previous section (so that each tetrahedron in the interior of the manifold is labelled twice). In \cite{4}, a basis for the space of vertices was determined by specifying a labelling of the tetrahedron by a spin $j$. In the new model, each basis vertex will be a certain linear combination of the $\vcenter{\epsfbox{canonical.eps}}$ labelled on the middle edge with $(j,j)$. 

There are two slightly different possibilities here. The first is to 
label tetrahedra in each 4-simplex with vertices as just described. The 
second possibility is to label the tetrahedron in the manifold with $\vcenter{\epsfbox{canonical.eps}}$
labelled on the middle edge with $(j,j)$ and use this in each of the two 
4-simplexes it shares. The extra label $j$ is then summed with the same 
weights as given by the vertex. The second sum is thus the `diagonal'
subset of the first. Thus we actually have two slightly different models, 
both of which deserve careful study. If vertices other than the canonical 
vertex exist, there will be still other variant state sums to consider. 

In either case, a weight (in $\C$) is defined for the labelling, and the state sum model determines a number by summing the weights over all labellings.

For each 4-simplex, a weight is determined by taking the product of all the vertices for its tetrahedra to form a closed relativistic spin network. The weight for the manifold is determined by taking the product of all these weights with correction factors for the simplexes of dimension less than four. For example, on a tetrahedron, this is the inner product between the basis elements in the vector spaces of vertices for the tetrahedron in each of the two 4-simplexes it shares.
 This ensures that the state sum does not depend on the choice of basis for the space of vertices on tetrahedra in the interior. The lower dimensional corrections would be analogous to the ones in \cite{4}, but modified to take into account the fact that we are quantum tracing
over two copies of each representation at each step.

In \cite{4} we combined the 15 spins on each 4-simplex into a diagram
called a 15j-q symbol which depended on the orientation of the 4-simplex.
Since reversing orientation switches self-dual with anti-self-dual
bivectors, in the new state sum the weights are linear combinations of
 the product of two 15j-q symbols with identical representations but opposite orientation for each 4-simplex of the triangulated 4-manifold.  Thus the model is a `constrained' doubling of the Crane-Yetter model.

\head V. Conclusions\endhead
The one issue which still needs to be addressed in order
for the state sum we are proposing to be a candidate for a quantum
theory of gravity is how probability amplitudes computed with it
behave as we refine the triangulation. This is analogous to the
ultraviolet problems long familiar to quantum field theory and quantum
gravity. It is important to note however that the nature of the
problem has changed, since we no longer have a continuum metric.
Instead of short distance behaviour, we are concerned with fluctuations
within fluctuations, with each fluctuation having a characteristic
Planck length `size'. This at any rate is the most natural
interpretation of the labellings in our state sum.

There are several possible avenues for constructing a finite theory
from our state sum. The most elegant solution would be if there is topological invariance. It will take a good deal of careful calculation of
relativistic spin networks to determine if this is the case.

Another possibility is to consider the possibility that below a
certain subdivision the universe is in a self-dual, ie topological
state. However this would require relaxing the constraints to allow self-dual states in the model. We could think of this hypothesis as
related to the idea that the universe began in a topological
state.\cite{28}
 
Another possibility is to try to imitate our construction using either
the category of representations of a larger quantum group or possibly
a quantum supergroup. It is plausible that the sort of cancellations
familiar from supergravity and supersymmetry will reappear in our
context, and give us a finite theory. The advantage of this line of
development over superstring theory is that if the cancellations do
give a finite theory, it will be an exact theory rather than a
perturbative expansion to something more mysterious.

It is interesting that the sort of tensor categories which go into the
state sum we are proposing are so similar to the ones invented in constructing
string theories \cite{29}. Terms in our state sum can be interpreted as diagrams
in string perturbation theory, by connecting together the diagrams we
are associating to the 4-simplexes, and interpreting elements in the
representations as string states.

At this point, perhaps the strongest thing which we can say of our
model is that it transforms a number of important and familiar
questions into statements about finite calculations which are well
defined and can be investigated. We believe that this is a good sign
in itself.


\Refs 

\ref \no1 \by L. Crane \paper Clock and category, is Quantum Gravity Algebraic? \jour J. Math. Phys \vol36 \pages 6180--6193 \yr 1995
\endref
 
\ref \no2 \by L. Crane \paper Topological Field Theory as the Key to Quantum Gravity \inbook
Knots and Quantum gravity  \ed J.C. Baez  \yr 1994\publ Clarendon \publaddr Oxford\endref

\ref \no3 \by J. W. Barrett \paper Quantum Gravity as Topological Quantum Field Theory \jour J. Math. Phys. \vol 36 \pages 6161--6179\yr 1995\endref

\ref\no4 \by L. Crane, L. Kauffman and D. Yetter \paper State Sum invariants of
4-Manifolds \jour J. Knot Theory Ram. \vol 6  \yr 1997 \pages 177--234\endref

\ref\no5 \by H. Ooguri \paper Topological Lattice Models in Four Dimensions  Hirosi Ooguri \jour Mod. Phys. Lett. \vol A7 \yr 1992 \pages 2799--2810\endref

\ref \no6 \by J.C. Baez \paper Four-Dimensional BF Theory as a Topological Quantum Field Theory \jour Lett. Math. Phys. \vol 38 \yr 1996 \pages 129--143 \paperinfo q-alg/\discretionary{}{}{}9507006\endref

\ref\no7 \by J. Roberts \paper Skein theory and Turaev-Viro invariants \jour Topology \pages 771--788 \vol 34 \yr 1995\endref

\ref\no8 \by J.F. Pleba\'nski \paper On the separation of Einsteinian substructures \jour J. Math. Phys. \vol 18 \pages 2511--2520 \yr 1977\endref

\ref\no9 \by R. Capovilla, J. Dell, T. Jacobson, L. Mason \paper  Self-dual 2-forms and gravity \jour Class. Quantum Grav. \vol 8 \yr 1991 \pages 41--58\endref

\ref\no10 \by G.T. Horowitz \paper Exactly soluble diffeomorphism invariant theories \jour Comm. Math. Phys. \vol 125 \pages 417--437 \yr 1989 \endref

\ref\no11 \by L. Smolin \paper Linking Topological Quantum Field Theory and
Nonperturbative Quantum Gravity \paperinfo gr-qc/9505028 \jour J. Math. Phys. \vol 36 \yr 1995 \pages 6417--6455\endref

\ref\no12 \by G. Ponzano and T. Regge \paper Semiclassical limit of Racah
coefficients \inbook Spectroscopic and group theoretical methods in
Physics \ed F. Bloch \publ North-Holland \yr 1968\pages 1--58
\endref

\ref \no13 \by E. Witten \paper 2+1 Gravity as an Exactly Soluble System \jour Nucl. Phys.
\vol B311 \pages 46--78 \yr 1988
\endref

\ref \no14 \by C. Rovelli and L. Smolin \paper Discreteness of Area and 
Volume in Quantum Gravity \jour Nucl. Phys.  \vol B442 \yr 1995 \pages 593--622\endref

\ref\no15 \paper The basis of the Ponzano-Regge-Turaev-Viro-Ooguri model is the loop representation basis \by C. Rovelli \jour Phys.Rev. \vol D48 \yr 1993 \pages 2702--2707 \endref

\ref\no16 \paper Sum over Surfaces form of Loop Quantum Gravity
      \by M.P. Reisenberger, C. Rovelli
\jour Phys.Rev. \vol D56 \yr 1997 \pages 3490--3508\endref

 \ref\no17  \paper A Hamiltonian lattice formulation of topological gravity
        \by H. Waelbroeck and J.A. Zapata \yr 1994 \jour Class. Quantum Grav. \vol 11 \pages 989--998 \endref

\ref\no18 \paper Topological lattice gravity using self-dual variables 
     \by   J.A. Zapata \yr 1996 \jour Class. Quantum Grav. \vol 13 \pages 2617--2634 \endref 

\ref\no19 \paper A left-handed simplicial action for Euclidean general relativity \by M.P. Reisenberger \yr 1997 \jour Class. Quantum Grav. \vol 14 \pages 1753--1770\endref 

\ref \no 20 \by J.W. Barrett \paper A convergence result for linearised Regge calculus \jour Class. Quant. Grav. \vol 5 \yr 1988 \pages 1187--1192\endref

\ref \no 21 \by J.W. Barrett and R.M. Williams \paper The convergence of lattice solutions of linearised Regge calculus \jour Class. Quant. Grav. \vol 5 \yr 1988 \pages 1543--1556\endref

\ref \no22 \by J.C. Baez \paper Spin Foam Models \paperinfo gr-qc/9709052 \endref

\ref\no23 \by A. Barbieri  \paperinfo  gr-qc/9707010 \paper Quantum tetrahedra and simplicial spin networks\endref

\ref \no24 \by J.W. Barrett \paper First order Regge calculus \jour Class. Quant. Grav. \vol 11 \yr 1994 \pages 2723--2730\endref

\ref \no25 \by J.W. Barrett, M. Rocek, R.M. Williams \paper A note on area variables in Regge calculus \paperinfo gr-qc/9710056 \endref

\ref\no26 \by L.H. Kauffman \yr 1990\paper Spin networks and knot polynomials \jour Int. J. Modern Physics A \vol 5\pages 93--115\endref

\ref\no27 \by J. Lewandowski \paper Volume and quantizations \jour Class. Quantum Grav. \vol 14 \yr 1997 \pages 71--76 \endref

\ref\no28 \by E. Witten \paper Topological quantum field theory
\jour Comm. Math. Phys.
\vol 117 \pages 353--386 \yr 1988\endref

\ref \no29 \by G. Moore and N. Seiberg \paper Classical and Quantum Conformal Field Theory \jour Comm. Math. Phys. \vol 123 \pages 177--254 \yr 1989\endref


\endRefs
\enddocument